# Corrections to the Predictions for Atmospheric Neutrino Observations

J. Poirier

*Physics Department, 225 NSH, University of Notre Dame, Notre Dame, IN 46556, USA*

## Abstract

The theoretical Monte Carlo (MC) calculations of the production of neutrinos via cosmic rays incident upon the earth's atmosphere are examined as well as their comparison to the experimental results. The MC calculations are sensitive to the assumed ratios of $\pi^+/\pi^-$ production cross sections; this ratio appears to be underestimated in the theory relative to the experimentally measured ratio. The MC calculation for the ratio $R = (\nu_\mu + \bar{\nu}_\mu)/(\nu_e + \bar{\nu}_e)$ thus has too few $\nu_e$'s in the denominator relative to the $\bar{\nu}_e$ as a direct result of a deficient $\pi^+/\pi^-$ ratio. Since the neutrino detection cross section is larger than that for the antineutrino, the theoretically predicted *detected* ratio, R', is artificially too large. By using a correct production ratio for $\pi^+/\pi^-$, the MC calculation yields a lower value for R', more in agreement with the experimental value. Correcting the theoretical value of R' shrinks it by a factor of 0.77. Thus the need for a physics explanation for the difference (like the neutrino oscillation hypothesis) is reduced and the likelihood of a statistical fluctuation is enhanced. Other corrections to the theory (see for example Poirier, 1999) could lower the theoretical value of R' still further, perhaps eliminating the difference between theory and experiment entirely.

## 1   Introduction:

The theoretical Monte Carlo (MC) calculations of the production of neutrinos via cosmic rays incident upon the earth's atmosphere have been addressed by several authors (Barr, Gaisser, & Stanev, 1989; Becker-Szendy et al., 1992; Bugaev & Naumov, 1989; Gaisser, Stanev, & Barr, 1988; and Honda et al., 1995). These calculations have been compared to experiments like Super-Kamiokande (Fukuda, 1998). The experiments measure a lower value for the *detected* ratio $R' = (\nu_\mu + \bar{\nu}_\mu)/(\nu_e + \bar{\nu}_e)$ than that predicted by the MC calculations. The Super-Kamiokande group has suggested a neutrino oscillation hypothesis to explain their lower observed value for R' by oscillating the $\nu_\mu$'s away, thus reducing the numerator of R'. It is here suggested that the value of R' is rather reduced by correcting the MC theory by using corrected input values to the calculations. This increases the denominator of the R' ratio, thus decreasing R' in a simpler manner.

The ratio of $\nu_e/\bar{\nu}_e$ (an output of the MC calculation) is a direct measure of the assumed $\pi^+/\pi^-$ production cross sections which parented these neutrinos through the $\pi \rightarrow \mu \rightarrow \nu_e$ decay chain. This can be seen from the reactions in Table 1 which show that a $\pi^+$ parent yields one $\nu_e$ whereas the $\pi^-$ yields one $\bar{\nu}_e$. This MC ratio is 1.16 for $\nu_e/\bar{\nu}_e$ at 1 GeV, which becomes a statement about their input values for the $\pi^+/\pi^-$ production cross section ratio. This ratio appears to be an underestimate of experimentally measured ratios which give a value of 3.9 at a relevant pion parent energy. Since the $\nu_e$ from $\pi^+$ decay chain is three times more likely to be detected than its antiparticle from the $\pi^-$ decay chain, the detected ratio $(\nu_\mu + \bar{\nu}_\mu)/(\nu_e + \bar{\nu}_e)$ has a denominator sensitive to the excess of $\nu_e$ over $\bar{\nu}_e$, an excess not properly modelled in the MC calculations. The net effect reduces the MC calculation for this ratio by a factor of 0.77.

The observed ratio of ratios [R'(experiment) / R'(theory)] is a statement that, relative to the MC calculation's value for R', the experimental values are R' are 0.63 and 0.65 in two different energy regions (Fukuda et al., 1998). By correcting the theory for this one effect alone, the difference between MC theory and experiment shrinks significantly. Other possibilities exist as well (see for example, Poirier, 1999, which would affect this overall ratio, as well as Super-Kamiokande's up / down ratio). Thus the necessity for a neutrino oscillation hypothesis is lessened as the possibility of a statistical fluctuation is enhanced or additional corrections to theory are identified and quantified.

## 2  Input values for $\pi^+/\pi^-$ production ratios in the MC calculation:

Table 1 gives the decay chain $\pi$ to $\mu$ to e for the $\pi^+$ and the $\pi^-$ decays and the relevant interaction cross sections for the neutrinos.

```
                                                              detection
         decay chain                 ν_μ   ν̄_μ   ν_e   ν̄_e   ratio, R'
π⁺  →   ν_μ  (μ⁺ →  e⁺   ν_e  ν̄_μ)    3    1     3            1.33
π⁻  →   ν̄_μ  (μ⁻ →  e⁻   ν̄_e  ν_μ)    3    1           1      4
         1:1    totals:               8          4             2
         3.36:1 totals:              17.4       11.1           1.57
```

**Table 1:** Decay chains for the $\pi^+$ and the $\pi^-$ with the cross section for the detection of the respective neutrinos (ν) relative to that for the antineutrino (ν̄) species. In the detection ratio R' = $(\nu_\mu + \bar\nu_\mu)/(\nu_e + \bar\nu_e)$, notice the numerator is the same between the $\pi^+$ and the $\pi^-$ whereas there is a difference of a factor of 3 in the denominator; only an equal mixture of $\pi^+$ and $\pi^-$ parents gives the ratio 2. The bottom line gives the corrected ratio for a revised $\pi^+/\pi^-$ ratio; R' = 1.57, or, relative to its expected value of 2, a reduction of 0.87.

The MC calculations are sensitive to the assumed production cross section the $\pi^+$ to $\pi^-$ ratio. As for the electron-type neutrinos, it is seen that $\pi^+$ gives only $\nu_e$ whereas the $\pi^-$ gives only the $\bar\nu_e$. Thus the $\pi^+/\pi^-$ ratio used in the MC calculations can be examined by looking at the ratio of $\nu_e / \bar\nu_e$ output of the MC calculations; this ratio is the $\pi^+/\pi^-$ production ratio used in the MC [at an appropriate higher pion energy (see section 3 below)]. For comparison, take a neutrino energy of 1 GeV. The various MC give the $\nu_e / \bar\nu_e$ ratios at the location of Super-Kamiokande of: 1.16 (Gaisser) 1.16 (Barr); 1.15 (Honda); and 1.16 (Bugaev). Due to the one-to-one correspondence between π+ and $\nu_e$ ($\pi^-$ and $\bar\nu_e$), it follows that the cross sections ratios for the production of $\pi^+/\pi^-$ used in these calculations are 1.16 ± .01 for those pions which parented the 1.0 GeV neutrinos.

## 3  Experimentally measured $\pi^+$ and $\pi^-$ production cross sections:

The experimental value for the $\pi^+/\pi^-$ ratio depends on the energy of the proton beam, so an appropriate energy must be selected for the pion parent to yield the 1.0 GeV $\nu_e$ in the example in section 2 above.

```
         experiment     1     2     3    3.7 GeV/c p momentum; References:
         9.75  deg:    2.4   2.6   2.9    ---         (Bevatron, 1971)
         12    deg:    2.1   2.6   4.1    ---         (Bevatron, 1971)
         16    deg:    2.2   2.6   ---    ---         (Bevatron, 1971)
         Hart+         ---   ---   ---    5.3         (Hart, Willis, 1962)
         Melissinos+   ---   ---   ---    4.3         (Melissinos, 1961)
         average       2.2   2.6   3.5    4.8
```

**Table 2:** Some experimental measurements of the $\pi^+/\pi^-$ production ratios from p + p interactions at proton beam momenta from 1.0 to 3.7 GeV/c.

The neutrino energy in the muon decay is approximately one-third that of the muon since there are three light particles in its final decay state. Therefore, a ~1.0 GeV neutrino would come from a ~3.0 GeV muon.

In the pion's decay to a muon, the muon takes most of the energy because of its large mass relative to the neutrino; a ~4 GeV pion would yield a most-likely-value of 3 GeV for its decay muon (at 90 degrees in its center of mass). Some experimental data in this energy region that relate to the ratio of $\pi^+ / \pi^-$ production cross sections are given in Table 2 above.

The highest value in the table is at the 3.7 GeV/c proton momentum. Take its average value of 4.8 for the experimental measurement of the $\pi^+ / \pi^-$ production ratio at this momentum, even though a higher momentum is needed and the ratio is rising. The target for the cosmic rays is primarily nitrogen, an even mixture of protons and neutrons; the $\pi^+$ to $\pi^-$ production ratio would not be as great as for a proton target. The p + p data has two positive charges in the initial state and therefore the final state must have an excess of two positive charges (charge conservation). For the case of p + n (neutron), there is an unbalance of only a single charge in the initial (and final) states. Therefore one might approximate the excess of $\pi^+$ for p + n to be half as large as the p + p case (because the excess charge is half as great). Using this approximation and averaging together the p + p with the p + n (to approximate the equal mixture of p + n in the nitrogen nucleus), then the $\pi^+ / \pi^-$ production ratio for p + N (N=nitrogen) would be 3.9 at the 3.7 GeV/c incident proton momentum. Since cosmic rays contain many components with excess positive charge *greater* than the proton, the estimate of a factor of 3.9 for the $\pi^+ / \pi^-$ production ratio is expected to be an underestimate of the overall $\pi^+ / \pi^-$ production ratio expected from the positive charge excess in the initial state of cosmic ray interactions. As well, measurements are needed at higher proton beam momentum and the ratio is rising. Nevertheless, the factor of 3.9 is taken as a conservative estimate for the measured $\pi^+$ to $\pi^-$ production ratio.

## 4 Experimental Neutrino and Antineutrino Interaction Cross Sections:

The Particle Data Group has compiled the neutrino and antineutrino interaction cross sections as a function of their laboratory energy (Aguilar-Benitez et al., 1986; the references for the experimental data can be found here) from 0 to 250 GeV. Taking the lowest energy points and taking into account the linear energy dependence of the interaction cross section, the neutrino cross section is about a factor of 3.0 higher than the antineutrino cross section. There is a scarcity of data at the relevant energy region for these experiments.

## 5 The Revised Detection Ratio, R', using the Proper $\pi^+ / \pi^-$ Ratio:

The interaction cross section for neutrinos is linear in the laboratory energy of the neutrinos. This causes an additional effect on the prediction for the detection ratio of the muon and electron neutrinos since the $\nu_\mu$ from the pion decay actually has a lower momentum than that for the $\nu_\mu$ from muon decay. In either case, the electron neutrino or antineutrino has the same energy on the average. As seen in section 3 above, for the example of a 1.0 GeV $\nu_e$, the accompanying $\nu_\mu$ from the muon decay is also 1.0 GeV, and the muon parent has 3 GeV of energy. The energy of the parent pion is 4.0 GeV and the $\nu_\mu$ from pion decay is 0.855 GeV energy (the latter two quantities are based on the most probable value which occurs at the 90 degree center of mass scattering angle). Notice that the $\nu_\mu$ energy from pion decay is *lower* than the 1.0 GeV $\nu_\mu$ from muon decay. Table 3 below is a repeat of Table 1 above, except now the muon-neutrino interaction cross sections have been modified to take into account their average energy dependence. The cross sections have been increased (reduced) so that the average effect keeps R' = 2. The 3.36:1 ratio of $\pi^+ / \pi^-$ is used to correct the value in the MC program (1.16) upward by the factor 3.36 in order to match the measured value of 4.8 (section 3). For example, the $\nu_\mu$ from $\pi^+$ decay is at a slightly lower-than-average energy, so its cross section is reduced from 3.00 to 2.68, etc.

|  decay chain | $\nu_\mu$ | $\bar{\nu}_\mu$ | $\nu_e$ | $\bar{\nu}_e$ | detection ratio, R' |
|---|---|---|---|---|---|
| $\pi^+ \rightarrow \nu_\mu \ (\mu^+ \rightarrow e^+ \ \nu_e \ \bar{\nu}_\mu)$ | 2.68 | 1.11 | 3 |  | 1.26 |
| $\pi^- \rightarrow \bar{\nu}_\mu \ (\mu^- \rightarrow e^- \ \bar{\nu}_e \ \nu_\mu)$ | 3.32 | .89 |  | 1 | 4.21 |
| 1:1 totals: | 8 | | 4 | | 2.00 |
| 3.36:1 totals: | 7.76 | | 5.08 | | 1.53 |

**Table 3:** The pion decay chains with the cross section for the detection of the respective neutrinos relative to the antineutrino. Table 1 is here modified for the neutrino interaction cross section energy dependence. The cross sections are adjusted so: their average is the same and still relative to the $\bar{\nu}_e$ cross section. A 50:50 mixture of $\pi^+$ and $\pi^-$ still gives the standard ratio of 2.0 as shown by the "1:1 totals" entry. The "3.36:1 totals" take into account the ratio of the experimentally measured $\pi^+ / \pi^-$ value of 3.9 versus the value used in the MC calculations of 1.16. R' is reduced from 2.00 to 1.53, a reduction factor of 0.77.

In summary, the calculation for the detected ratio R' = $(\nu_\mu + \bar{\nu}_\mu) / (\nu_e + \bar{\nu}_e)$ depends quite sensitively upon the parent $\pi^+$ to $\pi^-$ ratio; for the experimentally measured value of 3.9, R' has been reduced by a factor of 0.77 relative to the value obtained in the MC calculations for the assumed $\pi^+ / \pi^-$ ratio of 1.16. The statistical significance of any remaining discrepancy between theory and experiment is now considerably reduced.

## 6  Conclusion:

The effect discussed here considerably reduces the discrepancy between the MC calculations and the experimental data (see for example, Fukuda, et al., 1998) by lowering the MC calculation prediction for R' by a factor of 0.77, whereas the experimental factor is 0.63 and 0.65 for sub- and multi-GeV energies. Perhaps several such effects could bridge the remaining gap (for an example, see Poirier, 1999); or the effect which is only estimated here could, in fact, be larger (no increase was assumed in the $\pi^+ / \pi^-$ ratio for the fact that cosmic rays are more charged than +1 the $\pi^+ / \pi^-$ charge ratio measured at a higher momentum); or the uncertainty in the interaction cross sections for the neutrinos could be important; or it could be a statistical fluctuation of a smaller magnitude (some, or all, of the above). The necessity for a neutrino oscillation hypothesis is thereby lessened. Additional data on Super-Kamiokande's part are awaited along with enhancements to the MC calculations. Note added: Recent data (Messier, 1999) raises the sub-GeV measurement to 0.67.

Acknowledged are helpful discussions with John LoSecco who noted the difference between $\nu$ and $\bar{\nu}$ detection cross sections and its connection to the difference between the $\pi^+$ and $\pi^-$ production cross sections. Technical assistance from W. J. Carpenter, T. F. Lin, and A. Roesch was helpful.

This research is presently being funded through grants from the University of Notre Dame and private donations.